\documentstyle[preprint,prd,aps]{revtex}
\global\arraycolsep=2pt
\makeatletter
\def\fmslash{\@ifnextchar[{\fmsl@sh}{\fmsl@sh[0mu]}}
\def\fmsl@sh[#1]#2{%
  \mathchoice
    {\@fmsl@sh\displaystyle{#1}{#2}}%
    {\@fmsl@sh\textstyle{#1}{#2}}%
    {\@fmsl@sh\scriptstyle{#1}{#2}}%
    {\@fmsl@sh\scriptscriptstyle{#1}{#2}}}
\def\@fmsl@sh#1#2#3{\m@th\ooalign{$\hfil#1\mkern#2/\hfil$\crcr$#1#3$}}
\makeatother

\begin{document}
\draft\pagenumbering{roma}
\author{Yuan-Ben Dai$^a$, Chao-Shang Huang$^a$, Ming-Qiu Huang$^{b,c}$, 
Hong-Ying Jin$^c$, Chun Liu$^d$ }
\address{$^a$ Institute of Theoretical Physics, Academia Sinica, 
P.O.Box 2735, Beijing 100080, China}
\address{$^b$ CCAST (World Laboratory) P.O. Box 8730, Beijing, 100080}
\address{$^c$ Institute of High Energy 
Physics, Academia Sinica, P.O.Box 918, Beijing 100039, China}
\address{$^d$ Center For Theoretical Physics, Seoul 
National University, Seoul, 151-742, Korea}
\title{ Decay Widths of Excited Heavy Mesons From QCD Sum Rules
 at the Leading Order of HQET  }
\maketitle
\thispagestyle{empty}
\vspace{15mm}
\begin{abstract}
 The widths for pionic decays of lowest two excited doublets $(0^+,1^+)$ and $(1^+,2^+)$
of heavy mesons are studied with QCD sum rules in the leading order of heavy quark effective
theory. The ambiguity due to presence of two distinct $1^+$ states are solved.
The soft pion approximation is not used. 
\end{abstract}
\vspace{4mm}
 
\vspace{3.cm}
\newpage
\pagenumbering{arabic}
 
\section{Introduction}
\label{sec1} 

The properties of excited heavy mesons have attracted attention in recent
years. The mass and decay widths have been studied with potential model
combining with data for $K$ mesons in \cite{eichten}, with relativistic
Bethe-Salpeter equation in \cite{dai1,dai2} and 
 with QCD sum rules in \cite
{colangelo}. In work \cite{huang} the masses of low lying excited heavy
mesons $(2^+,1^+)$ and $(1^+,0^+)$ were studied with QCD sum rules in the
heavy quark effective theory (HQET)\cite{grinstein} up to the order of $%
{\cal O}(1/m_Q)$, where $m_Q$ is the heavy quark mass. The present work is
devoted to the decay widths of these mesons in the same formalism.

One problem encountered in studying the decay widths of excited heavy mesons
with QCD sum rules is the following. Except for the lowest states $0^-$, $%
1^- $, the spectra contains a pair of states for any spin-parity $j^P$ with
close values in their masses but quite different in magnitudes of their
decay widths. In our case, one of the two $1^+$ states is a narrow resonance
decaying mainly by emitting a $D$ wave pion, while the other one is a very
wide resonance decaying by emitting a $S$ wave pion. An interpolating
current used for the narrow $1^+$ state with a small coupling to the other $%
1^+$ state may cause sizable error in the result of calculation. It is only
in the $m_Q\to\infty$ limit, there is a conserved quantum number $j_{\ell}$,
the angular momentum of the light component, which can be used to
differentiate the two states. Therefore, HQET has important advantage for
this purpose.

The proper interpolating current $J_{j,P,j_{\ell}}^{\alpha_1\cdots\alpha_j}$
for the states with the quantum number $j$, $P$, $j_{\ell}$ in HQET was
given in \cite{huang}. They were proved to satisfy the following conditions 
\begin{eqnarray}
\label{decay}
\langle 0|J_{j,P,j_{\ell}}^{\alpha_1\cdots\alpha_j}(0)|j',P',j_{\ell}^{'}\rangle&=&
f_{Pj_l}\delta_{jj'}
\delta_{PP'}\delta_{j_{\ell}j_{\ell}^{'}}\eta^{\alpha_1\cdots\alpha_j}\;,\\
\label{corr}
i\:\langle 0|T\left (J_{j,P,j_{\ell}}^{\alpha_1\cdots\alpha_j}(x)J_{j',P',j_{\ell}'}^{\dag
\beta_1\cdots\beta_{j'}}(0)\right )|0\rangle&=&\delta_{jj'}\delta_{PP'}\delta_{j_{\ell}j_{\ell}'}
(-1)^j\:{\cal S}\:g_t^{\alpha_1\beta_1}\cdots g_t^{\alpha_j\beta_j}\nonumber\\[2mm]&&\times\:
\int \,dt\delta(x-vt)\:\Pi_{P,j_{\ell}}(x)
\end{eqnarray}
in the $m_Q\to\infty$ limit, where $\eta^{\alpha_1\cdots\alpha_j}$ is the
polarization tensor for the spin $j$ state, $v$ is the velocity of the heavy
quark, $g_t^{\alpha\beta}=g^{\alpha\beta}-v^{\alpha}v^{\beta}$ is the
transverse metric tensor, ${\cal S}$ denotes symmetrizing the indices and
subtracting the trace terms separately in the sets $(\alpha_1\cdots\alpha_j)$
and $(\beta_1\cdots\beta_{j})$, $f_{P,j_{\ell}}$ and $\Pi_{P,j_{\ell}}$ are
a constant and a function of $x$ respectively which depend only on $P$ and $%
j_{\ell}$. Because of equations (\ref{decay}) and (\ref{corr}), the sum rule
in HQET for decay widths derived from a correlator containing such currents
receive no contribution from the unwanted states with the same spin-parity
as the states under consideration in the $m_Q\to\infty$. Starting from the
calculations in the leading order, the decay amplitudes for finite $m_Q$ can
be calculated unambiguously order by order in the $1/m_Q$ expansion in HQET.


\section{ Sum rules for decay amplitudes}

In the present work we shall confine ourselves to the lowest lying excited
states in the leading order of $1/m_Q$ expansion. Denote the doublet $%
(1^+,2^+)$ with $j_{\ell}=3/2$ by $(B_1,B_2^*)$ and the doublet $(0^+,1^+)$
with $j_{\ell}=1/2$ by $(B^{\prime}_0,B^{\prime}_1)$. From covariance and
conservation of the angular momentum of the light component in the $%
m_Q\to\infty$ limit, the amplitudes for the decay of $B_1$, $B_2^*$ to the
ground states $B$, $B^*$ in the doublet $(0^-,1^-)$ with $j_{\ell}=1/2$ have
the following forms 
\begin{eqnarray}
\label{coup1}
 M(B_1\to B^*\pi)&=&I\;\epsilon^*_{\mu}\eta_{\nu}(q_t^{\mu}q_t^{\nu}-\frac{1}{3}g_t^{\mu\nu}
 q_t^2)g(B_1,B^*)\;,\\
\label{coup2}
 M(B_2^*\to B\pi)&=&I\;\eta_{\mu\nu}q_t^{\mu}q_t^{\nu}g(B_2^*,B)\;,\\
 \label{coup3}
 M(B_2^*\to B^*\pi)&=&I\;i\varepsilon_{\alpha\beta\sigma\rho}\;\epsilon^{*\alpha}v^{\beta}
 \eta^{\sigma\mu}q_t^{\rho}q_{t\mu}g(B_2^*,B^*)\;,
\end{eqnarray} 
where $\eta_{\mu\nu}$, $\eta_{\mu}$ and $\epsilon_{\mu}$ are polarization
tensors for states $2^+$, $1^+$ and $1^-$ respectively. $q_{t\mu}=q_{\mu}-v%
\cdot qv_{\mu}$. $I=\sqrt{2}$, $1$ for charged and neutral pion
respectively. It can be shown \cite{falk} by combining heavy quark symmetry
and chiral symmetry that the constants in (\ref{coup1})-(\ref{coup3})
satisfy 
\begin{eqnarray}
\label{ggg}
g(B_2^*,B)=g(B_2^*,B^*)=\sqrt{\frac{2}{3}}\;g(B_1,B^*)\equiv {\cal G}\;.
\end{eqnarray} 

For deriving the sum rules for the constants in the decay amplitudes (\ref
{coup1}), (\ref{coup2}) and (\ref{coup3}) we consider the correlators 
\begin{eqnarray}
\label{7a}
 \int d^4x\;e^{-ik\cdot x}\langle\pi(q)|T\left(J^{\beta}_{1,-,\frac{1}{2}}(0)
 J^{\dagger\alpha}_{1,+,\frac{3}{2}}(x)\right)|0\rangle\hspace{7.3cm}\nonumber\\
 =\int\int d^4 xd^4 y\;e^{-ik\cdot x+ik^\prime \cdot y}\langle 0|T
 \left(J^{\beta}_{1,-,\frac{1}{2}}(y)\frac{1}{f_{\pi}}\partial^{\mu}j^5_{\mu}(0)
 J^{\dagger\alpha}_{1,+,\frac{3}{2}}(x)\right)|0\rangle\hspace{.2cm}\nonumber\\=
 \left(q_t^{\alpha}q_t^{\beta}-\frac{1}{3}g_t^{\alpha\beta}q^2_t\right)G_{B_1B^*}
 (\omega,\omega')\;,\hspace{5.3cm}\\[2mm]\label{7b}
\int d^4x\;e^{-ik\cdot x}\langle\pi(q)|T\left(J_{0,-,\frac{1}{2}}(0)
 J^{\dagger\alpha\beta}_{2,+,\frac{3}{2}}(x)\right)|0\rangle=
 \left(q_t^{\alpha}q_t^{\beta}-\frac{1}{3}g_t^{\alpha\beta}q^2_t\right)G_{B_2^*B}
 (\omega,\omega')\;,\hspace{1.4cm}\\[2mm]\label{7c}
 \int d^4x\;e^{-ik\cdot x}\langle\pi(q)|T\left(J^{\gamma}_{1,-,\frac{1}{2}}(0)
 J^{\dagger\alpha\beta}_{2,+,\frac{3}{2}}(x)\right)|0\rangle
 ={i\over 2}\left(\varepsilon^{\rho\gamma\alpha\sigma}v_\rho q_{t\sigma}q_t^{\beta}+\alpha
 \leftrightarrow\beta\right)G_{B^*_2B^*}(\omega,\omega')\;,
\end{eqnarray}
where $k=k^{\prime}+q$, $\omega=2v\cdot k$, $\omega^{\prime}=2v\cdot
k^{\prime}$ and $q^2=0$. PCAC is used in the first equality of (\ref{7a}).
The forms of the right hand side of (\ref{7a}), (\ref{7b}) and (\ref{7c})
are determined by that $\alpha$, $\beta$ and $\gamma$ are transverse
indices, $x-y=vt$ on the heavy quark propagator and that the pion is in the $%
D$ wave state due to conservation of angular momentum of the light
component. The interpolationg currents are given in \cite{huang} as 
\begin{eqnarray}
\label{curr1}
&&J^{\dag\alpha}_{1,+,{3\over 2}}=\sqrt{\frac{3}{4}}\:\bar h_v\gamma^5(-i)\left(
{\cal D}_t^{\alpha}-\frac{1}{3}\gamma_t^{\alpha}\fmslash{\cal D}_t\right)q\;,\\
\label{curr2}
&&J^{\dag\alpha_1,\alpha_2}_{2,+,{3\over 2}}=\sqrt{\frac{1}{2}}\:\bar h_v
\frac{(-i)}{2}\left(\gamma_t^{\alpha_1}{\cal D}_t^{\alpha_2}+
\gamma_t^{\alpha_2}{\cal D}_t^{\alpha_1}-{2\over 3}g_t^{\alpha_1\alpha_2}
\fmslash{\cal D}_t\right)q\;,\\
\label{curr3}
&&J^{\dag\alpha}_{1,-,{1\over 2}}=\sqrt{\frac{1}{2}}\:\bar h_v\gamma_t^{\alpha}
q\;,\hspace{1.5cm} J^{\dag\alpha}_{0,-,{1\over 2}}=\sqrt{\frac{1}{2}}\:\bar h_v\gamma_5q\;,
\end{eqnarray}
where $h_v$ is the heavy quark field in HQET and $\gamma_{t\mu}=\gamma_%
\mu-v_\mu\fmslash v$.

Let us first consider the function $G_{B_1B^*}(\omega,\omega^{\prime})$ in (%
\ref{7a}). As a function of two variables, it has the following pole terms
from double dispersion relation 
\begin{eqnarray}
\label{pole}
{f_{-,{1\over 2}}f_{+,{3\over 2}}g(B_1B^*)\over (2\bar\Lambda_{-,{1\over 2}}
-\omega^{\prime})(2\bar\Lambda_{+,{3\over 2}}-\omega)}+{c\over 2\bar\Lambda_{-,{1\over 2}}
-\omega^{\prime}}+{c'\over 2\bar\Lambda_{+,{3\over 2}}-\omega}\;,
\end{eqnarray}
where $f_{P,j_\ell}$ are constants defined in (\ref{decay}), $%
\bar\Lambda_{P,j_\ell}=m_{P,j_\ell}-m_Q$. As explained in Section \ref{sec1}%
, only one state with $j^P=1^+$ contributes to (\ref{pole}) as the result of
equation (\ref{decay}). This would not be true if the last term in (\ref
{curr1}) is absent.

Usually, the soft-pion approximation $\omega=\omega^{\prime}$ is used in QCD
sum rules for decay amplitudes. This is not a good approximation since the
difference $\omega-\omega^{\prime}=2v\cdot q$ is about $1$ GeV in the real
decay process as can be seen from the observed spectra of $D$ system and the
calculation results of \cite{huang}. We instead put 
\begin{eqnarray}
\label{pole1}
\omega-\omega^{\prime}=2(\bar\Lambda_{+,{3\over 2}}-\bar\Lambda_{-,{1\over 2}})=2\Delta\;.
\end{eqnarray}
This is satisfied by the real decay process in the leading order of $1/m_Q$.
We then use dispersion relation of one variable to $G_{B_1B^*}(\omega,%
\omega^{\prime})$, the pole terms of which are 
\begin{eqnarray}
\label{pole2}
{f_{-,{1\over 2}}f_{+,{3\over 2}}g(B_1B^*)\over (2\bar\Lambda_{-,{1\over 2}}
-\omega^{\prime})^2}+{c+c'\over (2\bar\Lambda_{-,{1\over 2}}
-\omega^{\prime})} \;.
\end{eqnarray}

The free heavy quark propagator in HQET is 
$\displaystyle{\int d\tau \;\delta(x-v\tau)\;{1+\fmslash v\over 2}}$.
The interaction of the heavy quark with the gluon field $A$ in the leading order of
$1/m_Q$ expansion is $g \bar h v\cdot A h$. We shall use the
 Fock-Schwinger gauge $x_{\mu}A^{\mu}(x)=0$ throughout this paper. In this
gauge the full
propagator of the heavy quark $\langle 0|T(h(0)\bar h(x)) |0\rangle$ 
in the leading order of $1/m_Q$ is identical to the free one, because $v=x/ \tau$ on the heavy
quark line. Therfore, the correlator (\ref {7a}) has the form
\begin{eqnarray}
\label{ad} 
\int d\tau \; e^{i\omega\tau}\; \mbox{tr}\{{\Gamma \langle \pi(q)|D_tq(v\tau)
\bar q(0)|0\rangle}\} = ({q_t}^{\alpha}{q_t}^{\beta}-{g_t}^{\alpha\beta})
 \int d\tau \; e^{i\omega\tau}f(vq,\tau), 
\end{eqnarray}
where $\Gamma$ is a vertex factor containing $\gamma$ matrices. Since in our
approach $v\cdot q$ is fixed to be equal to $\Delta$ by (\ref{pole1}) and the
invaraint function $G(B_1B^*)$ is considered as a function of a single
variable $\omega$ (or $\omega^{\prime}$), the integral in $\tau$ is dominated 
by small $\tau$, that is, the small distance $x=v\cdot \tau$ for $\omega$ 
in the deep euclidian region. Therefore
for deriving QCD sum rules we can calculate the correlator (\ref{7a}) by the
operator product expansion. This is an expansion over the small quantity $\tau$, 
whch after the integration becomes an assymptotic
expansion in powers of $1/ \omega$. By dimensional reason it is a power series in 
$v\cdot q/\omega$ and  $\Lambda_{QCD}/\omega$ which is meaningful for sufficiently large value
of $\omega$. After Borel transform the series is convergent when $v\cdot q/T$ is 
sufficiently small, where T is the Borel transform parameter. We emphasize that this is so
because $v\cdot q$ is fixed in our approach.
The perturbation terms do not contribute since
they do not violate the conservation equation $\partial^\mu j^5_\mu=0$. For
the higher terms we expand the quark field $q(x)$ in the matrix element $%
\langle\pi(q)|T(q(x)\bar q(0))|0\rangle$. The matrix elements appearing in
the expansion have the general form (\ref{a1}), (\ref{a2}) and (\ref{a3}) in
Appendix. The terms in (\ref{a1})-(\ref{a3}) which contribute to the QCD sum
rules in the soft pion approximation were used in \cite{colangelo}. Here we
keep all terms in these formulas in the calculations in order to avoid the
soft pion approximation. Substituting these equations in the correlator (\ref{7a}) 
and making Borel transform to supress the contribution of higher
states we find the sum rule in the leading order of HQET. 
\begin{eqnarray}
\label{sum1}
f_{-,{1\over 2}}f_{+,{3\over 2}}{g(B_1B^*)\over T}e^{-2\bar\Lambda_{-,{1\over 2}}/T}
+(c+c')e^{-2\bar\Lambda_{-,{1\over 2}}/T} + \cdots=\hspace{1cm} \nonumber\\
\sqrt{{3\over 2}}[g_2+{2\over T}(c_2-d_2-e_2\Delta)]e^{2\Delta/T}\;.
\end{eqnarray}
In the above equation we have neglected all radiative corrections and
operators with dimension $\geq 6$.
In the derivation of above equation we need to rotate the integration path clock-wise to
the imaginary axis of $\tau$ and  redefine the integral variable $\tau'=i\tau$. 
The result is then obtained by using the Borel transform formula 
  $\displaystyle{ \hat{B}^\omega_{T} \int d\tau\, e^{i \tau \omega}%
    \tau^n =(-i)^{n+1} (1/T)^n}$
in the negative region of $\omega$.

The dots in the left hand side of (\ref{sum1}) represents terms from the contributions
from the radial excited states of the same $j^P$ and $j_\ell$ and the continuum.
In order to estimate these terms we make the duality assumption that above $%
\omega=\omega_c$ the dispersion integral of the hadron side is equal to 
that obtained from the QCD calculation. The spectra function $\rho_{\mbox{\small QCD}}%
(\omega)$ of the latter is obtained by performing the Borel transform 
$\hat B^T_{1/\omega}$ to the right hand side of (\ref{sum1}). The result
for $\rho_{\mbox{\small QCD}}(\omega)$ contains two terms proportional to 
$\delta(\omega-2\Delta)$ and $\delta'(\omega-2\Delta)$ respectively. This 
implies that $\displaystyle{\int^\infty_{\omega_c}{\rho_{\mbox{\small QCD}}(\omega')d\omega'\over%
\omega'-\omega}}$ vanishes for $\omega_c>2\Delta\sim 1$ GeV. Therefore we shall neglect
the contributions from the radial excited states and the continuum to the sum rule
(\ref{sum1}).

We can multiply (\ref{sum1}) by $exp(2\bar \Lambda _{-,{\frac 12}}/T)$ and
differentiate in $1/T$ in order to eliminate $c$ and $c^{\prime }$. By using (%
\ref{ggg}) we obtain the following sum rule for the constant ${\cal G}$ 
\begin{eqnarray}
\label{sum1a}
{\cal G}f_{-,{1\over 2}}f_{+,{3\over 2}}= 2\{ \bar\Lambda_{+,{3\over 2}}g_2+
(c_2-d_2-e_2\Delta)(1+{2\bar\Lambda_{+,{3\over 2}}\over T})\}
e^{2\bar\Lambda_{+,{3\over 2}}/T} \;.
\end{eqnarray}

The sum rules for the correlators (\ref{7b}) and (\ref{7c}) can be obtained by
similar calculations. Equations similar to (\ref{sum1a}) are obtained and we
have checked that the relations in ( \ref{ggg} ) are satisfied.

The decay amplitudes of the members of the doublet $(j,+,1/2)$, $B_0^{\prime
}$ and $B_1^{\prime }$, have the following form 
\begin{eqnarray}
\label{coup4}
 M(B'_0\to B\pi)&=&I\;{\cal G'} \;,\\
\label{coup5}
 M(B_1'\to B^*\pi)&=&I\;\epsilon^*\cdot\eta {\cal G'} \;
  \end{eqnarray}
in the $m_Q\to \infty $ limit. The sum rule in HQET for the coupling
constant ${\cal G^{\prime }}$ can be derived similarly. From \cite{huang},
the interpolating currents for $B_0^{\prime }$, $B_1^{\prime }$ can be
chosen as 
\begin{eqnarray}
\label{current1}
J^{\dag}_{0,+,{1\over 2}}=\frac{1}{\sqrt{2}}\:\bar h_vq\;,\hspace{1.5cm}
J^{\dag\alpha}_{1,+,{1\over 2}}=\frac{1}{\sqrt{2}}\:\bar h_v\gamma^5\gamma^{\alpha}_tq\;.
\end{eqnarray} 
As pointed out in \cite{huang}, for any state $(j,P,j_\ell )$ besides the
interpolating current with the lowest dimension there is another one which
has an additional factor $-i\fmslash{\cal D}_t$ inserting before $q$. For
the doublet $(j,+,1/2)$, they are 
\begin{eqnarray}
\label{current2}
J^{'\dag}_{0,+,{1\over 2}}=\frac{1}{\sqrt{2}}\:\bar h_v(-i)\fmslash{\cal D}_tq\;,\hspace{1.5cm}
J^{'\dag}_{1,+,{1\over 2}}=\frac{1}{\sqrt{2}}\:\bar h_v\gamma^5\gamma^{\alpha}_t(-i)\fmslash{\cal D}_tq\;.
\end{eqnarray}
As discussed in \cite{huang}, there is some motivation for using the latter
currents for the corresponding constant $f_{+,1/2}^{\prime }$ survives in
the non-relativistic limit of the Bethe-Salpeter wave function, while the
constant $f_{+,1/2}$ for the current (\ref{current1}) vanishes in this
limit, since $B_0^{\prime }$ and $B_1^{\prime }$ are $P$ wave states in the
non-relativistic quark model.

After Borel transformation and eliminating the single pole terms by
differentiation, the sum rule derived by using the interpolating current (%
\ref{current2}) is 
\begin{eqnarray}
\label{sum2}
f_{-,{1\over 2}}f'_{+,{1\over 2}}{\cal G'}&=&2[\bar\Lambda_{+,{1\over 2}}(-g_2\Delta^2+3h_2\Delta
+3h_1)-(-3b_1+3c_1\Delta+3d_1\Delta\nonumber\\&&\mbox{}+4c_2\Delta^2+2d_2\Delta^2-e_2\Delta^3)
(1+{2\bar\Lambda_{+,{1\over 2}}\over T})]
e^{2\bar\Lambda_{+,{1\over 2}}/T}\;,
\end{eqnarray}
where $\Delta=\bar\Lambda^{\prime}_{+,1/2}-\bar\Lambda_{-,1/2}$. If the interpolating
current (\ref{current1}) is used instead of (\ref{current2}), the sum rule
turns out to be 
\begin{eqnarray}
\label{sum3}
f_{-,{1\over 2}}f_{+,{1\over 2}}{\cal G'}=\left((-\bar\Lambda_{+,{1\over 2}}\alpha-{\beta\over 2})
-{\bar\Lambda_{+,{1\over 2}}\beta-\gamma\over T}+{\bar\Lambda_{+,{1\over 2}}\gamma\over T^2} \right)
e^{2\bar\Lambda_{+,{1\over 2}}/T}\;,
\end{eqnarray}
where 
\begin{eqnarray}
\label{para}
&&\alpha=2(f_1+f_2\Delta)\;,\hspace{1cm}\beta=4(g_1\Delta+g_2\Delta^2+h_1)\;,\nonumber\\
&&\gamma=4(a_1+a_2\Delta+2c_1\Delta+e_1\Delta^2+e_2\Delta^3)\;
\end{eqnarray}
and $\Delta=\bar\Lambda_{+,1/2}-\bar\Lambda_{-,1/2}$. 

\section{Determination of the parameters}
\label{sec3}
 
In order to obtain the decay constants ${\cal G}$ and ${\cal %
G^{\prime}}$ from (\ref{sum1a}), (\ref{sum2}) and (\ref{sum3}) we need to use
the mass parameters $\bar\Lambda$'s and the coupling constants $f$'s of the
corresponding interpolating currents as input. $\bar\Lambda_{-,1/2}$ and $%
f_{-,1/2}$ can be obtained from the results in \cite{neubert} as $%
\bar\Lambda_{-,1/2}=0.5$ GeV and $f_{-,1/2}\simeq 0.24$ GeV$^{3/2}$ at the
order $\alpha_s=0$. Notice that the coupling constant $f_{-,1/2}$ defined in
the present work is a factor $1/\sqrt 2$ smaller than that defined in \cite
{neubert}. $\bar\Lambda_{+,3/2}$, $\bar\Lambda_{+,1/2}$ and $%
\bar\Lambda^{\prime}_{+,1/2}$ are given in \cite{huang}. $f_{+,3/2}$, $%
f_{+,1/2}$ and $f^{\prime}_{+,1/2}$ can be determined from the formulas
(34), (27) and (28) of reference \cite{huang} derived from sum rules for two
point correlators. The results are 
\begin{eqnarray}
\label{fvalue}
&&\bar\Lambda_{+,3/2}=0.95 ~~\mbox{GeV}\hspace{1.2cm}f_{+,3/2}=0.19\pm 0.03 ~~\mbox{GeV}^{5/2}\;,\nonumber\\
&&\bar\Lambda'_{+,1/2}=0.9 ~~\mbox{GeV}\hspace{1.3cm}f'_{+,1/2}=0.28\pm 0.03 ~~\mbox{GeV}^{5/2}\;,\nonumber\\
&&\bar\Lambda_{+,1/2}=1.05 ~~\mbox{GeV}\hspace{1.1cm}f_{+,1/2}=0.32\pm 0.03 ~~\mbox{GeV}^{3/2}\;.
\end{eqnarray}

(\ref{a4})-(\ref{a64}) still contains several parameters. $m_0^2$ is known
to be $0.8$ GeV$^2$. $m_1^2$ was calculated in \cite{shifman} with QCD sum
rules with the result $m_1^2=0.2$ GeV$^2$. We have obtained QCD sum rules
for $e_1$ and $e_2$ by considering the two point functions 
\begin{eqnarray}
\label{e1}
&&i\int e^{iq\cdot x}d^4x\langle 0|T\{\bar q(x)i\gamma_5{\tau^a\over 2}q(x)
\bar q(0){\tau^b\over 2}(n\cdot D)^2i\gamma_5q(0)|0\rangle=(n\cdot q)^2A(q^2)\delta_{ab}\;,\\[3mm]
\label{e2}
&&i\int e^{iq\cdot x}d^4x\langle 0|T\{\bar q(x)\fmslash n\gamma_5{\tau^a\over 2}q(x)
\bar q(0){\tau^b\over 2}(n\cdot D)^2\fmslash n\gamma_5q(0)|0\rangle=
(n\cdot q)^4 B(q^2)\delta_{ab}\;,
\end{eqnarray}
where $n^2=0$. The resulting QCD sum rules are 
\begin{eqnarray}
\label{sume1}
{2e_1\over f_\pi}\langle\bar qq\rangle_0&=&{1\over 16\pi^2}\int_0^{s_0}e^{-{s/M^2}}ds
+{1\over 24\pi}\langle\alpha_s G^2\rangle_0+{112\over 81}\pi\alpha_s{\langle\bar qq\rangle_0^2
\over M^2}\;,\\[3mm]
\label{sume2}
2e_2f_\pi m_A^2&=&{3\over 80\pi^2}\int_0^{s_0}(m_A^2-s)e^{-{s/M^2}}ds+{1\over 48\pi}\langle\alpha_s G^2\rangle_0
\left(1+{m_A^2\over M^2}\right)\nonumber\\&&\mbox{}+{20\over 27}\pi\alpha_s\left(2+{m_A^2\over M^2}\right)
 {\langle\bar qq\rangle_0^2\over M^2}\;.
\end{eqnarray}
In obtaining (\ref{sume2}) we have eliminated the $a_1$ pole term by
differentiation with respect to the Borel parameter $1/M^2$. $m_{A}\simeq
1.3 $ GeV is the mass of $a_1$. In the numerical calculations for $e_1$ and $%
e_2$ we use the following standard values of the vacuum condensates: 
\begin{eqnarray}
\label{parameter}
\langle\bar qq\rangle_0&=&-(0.24 ~\mbox{GeV})^3\;,\nonumber\\
\langle\alpha_s GG\rangle_0&=&0.038 ~\mbox{GeV}^4\;
\end{eqnarray}
as well as $\alpha_s=0.4$. From (\ref{sume1}) and (\ref{sume2}) we obtain
the following numerical values for $e_1$ and $e_2$, 
\begin{eqnarray}
\label{e1e2}
&&e_1=-0.014\pm 0.002 ~~~\mbox{GeV}^2\;,\nonumber\\&&e_2=0.015\pm 0.002
~~~\mbox{GeV}\;.
\end{eqnarray}

\section{Numerical results and discussion}

\label{sec4}

We now turn to the numerical evaluation of the sum rules for ${\cal G}$ and $%
{\cal G^{\prime}}$. By using (\ref{fvalue}) and (\ref{parameter}) together
with the values of parameters determined above we obtain ${\cal G}$ and $%
{\cal G^{\prime}}$ as functions of $T$ from the sum rules (\ref{sum1a}), (\ref
{sum2}) and (\ref{sum3}). The results are plotted as curves in fig. 1. The
lower limit of $T$ is determined by the requirement that in the sum rules 
before differentiating with respect to
$1/T$ (such as (\ref{sum1})), the terms of higher
order in $1/T$  in the operator expansion are reasonably smaller than the
leading term, say $\leq 1/3$ of the latter. This leads to $T > 1.2$  GeV
and $ T > 3$ GeV for
the sum rule   (\ref{sum1a}) and (\ref{sum2}) respectively. 

 Usually the value of $T$ is constrained from the upper side by the requirement that
 the contribution from higher states is sufficiently suppressed. The continuum model
 discussed below equation (\ref{sum1}) allows $T$ to be any large value.
 However, in order that the results depend less on the continuum model it is good 
 if $T$ is not much larger than the $2(\bar \Lambda^{(1)}-\bar \Lambda )$,
 where $m_Q+ \bar \Lambda^{(1)}$ is the mass of the radial excited state. 
 Quark model calculations \cite{godfrey} indicate that $\bar \Lambda^{(1)}-\bar \Lambda$
 is roughly $0.7$ GeV. Therefore, this requirement can be satisfied for the sum rule for
 $\cal G$. We find within the stability window subject to the constrain
 disussed above
 \begin{eqnarray}
\label{result1}
 {\cal G}=5.7\pm 0.6\pm 0.7~~~\mbox{GeV}^{-2}\;, \hspace{1cm}1.3~~\mbox{GeV}
 <T<1.5~~\mbox{GeV}.
  \end{eqnarray}
 From the view point stated above, the sum rule (\ref{sum2}) for $\cal G'$ is
 not very satisfactory since the lower limit of $T$ is too high. If we somewhat
 arbitrarily take the width of the window to be $0.3$ GeV, we find from (\ref{sum2})
 \begin{eqnarray}
\label{result2}
 {\cal G'}=1.0\pm 0.1\pm 0.5\;, \hspace{1cm}  3~~\mbox{GeV}
 <T<3.3~~\mbox{GeV}.
  \end{eqnarray}
In (\ref{result1}) and (\ref{result2}) the  
 first error refers  to  the variation with $T$ in this region,
while the second error takes into account the uncertainty in the 
values of the coupling constant $f$'s. The errors quoted above do not
include those from higher order terms in OPE and the $1/m_Q$ expansion and the values
of parameters characterizing the condensations.

However,the (\ref{result1}) and (\ref{result2}) have another source of
uncertainty. These sum rules depend on the parameter $e_1$ which
is obtained by the sum rule (\ref{sume1}). It is well known \cite{shifman}
that the sum rules derived from the correlator of two pseudo-scalar
currents give too small value for the physical quantities. In the case of two
pseudo-scalar current without derivative, the phenomenogical side is about
a factor $2$ larger than the theoretical side. Just for a rough estimation,
let us multiplying the $e_1$ value by a factor $2$. We find that the $T$
value at which the last term is $1/3$ of the first term in the sum rule before 
differention with respect to $1/T$ move to lower values $0.6$ GeV and $1.9$ GeV
for the sum rule (\ref{sum1}) and (\ref{sum2}) respectively. That is,
the convergence becomes faster. The values of $\cal G$ and $\cal G'$ change to
\begin{eqnarray}
\label{result12}
 &&{\cal G}=3.6\pm 1.0\pm 0.7~~~\mbox{GeV}^{-2}\;, \hspace{1cm}1.0~~\mbox{GeV}
 <T<1.5~~\mbox{GeV}\;,\nonumber\\
 &&{\cal G'}=1.2\pm 0.1\pm 0.5\;, \hspace{2.6cm}  1.8~~\mbox{GeV}
 <T<2.1~~\mbox{GeV}\;.
  \end{eqnarray}
  
On the other hand the sum rule (\ref{sum3}) is not good in stability.
Especially, ${\cal G^{\prime}}$ changes sign near $T\simeq 1.2$ GeV.
Therefore, we do not use it in the numerical calculation of ${\cal G^{\prime}%
}$.

The decay widths in the leading order are 
\begin{eqnarray}
\label{widths}
&&\Gamma(B_1\to B^*\pi)={1\over 8\pi} {\cal G}^2|\vec q|^5\;,\nonumber\\
&&\Gamma(B_2^*\to B\pi)={1\over 20\pi} {\cal G}^2|\vec q|^5\;,\nonumber\\
&&\Gamma(B_2^*\to B^*\pi)={3\over 40\pi} {\cal G}^2|\vec q|^5\;,\nonumber\\
&&\Gamma(B'_0\to B\pi)={3\over 8\pi} {\cal G}^{'2}|\vec q|\;,\nonumber\\
&&\Gamma(B'_1\to B^*\pi)={3\over 8\pi} {\cal G}^{'2}|\vec q|\;,
\end{eqnarray}
where sum over charged and neutral pion final states has been included in
the above formulas.

In the following we apply the leading order formulas obtained above to the
excited states of charmed mesons. The value $\vec q$ for
the first three processes in (\ref{widths}) are calculated from the experimental
 mass values of the relevent particles \cite{review} and that for the last two
processes are calculated from the difference between the theoretical
mass values of the two doublets in the leading order obtained in \cite{huang}
 and \cite{neubert}. For the
excited states of $D$ we obtain the results listed in table \ref{tab:width}, where the
numbers in the parentheses are results for $e_1=-0.028$. The experimental
 data \cite{review} and the results of \cite{dai2} obtained with
relativistic B-S equation are also included. For the $D$ 
series the ${\cal O}(1/m_Q)$ corrections are not negligible. Yet, assuming the total 
width of 
$D^*_2$ is dominated by one pion decay, the calculation result for $D_2^*$
is in reasonable agreement with experimental data. The calculation result for 
$\Gamma(D_1\to D^*\pi)$ in the leading order is significantly smaller than the 
experimental data
for $\Gamma(D_1\to\mbox{all})$. This can partly be attributed to mixing of
the two $1^+$ states in the order of ${\cal O}(1/m_Q)$ \cite{falk}. This
effect can be calculated within our formalism by introducing the ${\cal O}%
(1/m_Q)$ terms in the Largrangian of HQET. In reference \cite{eichten}, it
was estimated that the $\Gamma(D_1\to D\rho+D^*\rho)\simeq 6$ MeV, $%
\Gamma(D_2^*\to D\rho+D^*\rho)\simeq 3$ MeV. Therefore, the effect of mixing
of two $1^+$ states in the order ${\cal O}(1/m_Q)$ needs not to be very
large. The calculated results for the decay widths of $D_0^{\prime}$
and $D_1^{\prime}$ in the leading order is too small. This may due to the slow
convergence of Eq. (\ref{sum2}).

\acknowledgments This work was supported in part by
the National Natural Science Foundation of China. One of the authers (C.L.)
was supported in part by the KOSEF through the SRC programm. 
\vspace{0.5cm}

{\large {\bf Appendix: The OPE of matrix element $\langle\pi^i(q)|T(q(x)\bar
q(0))|0\rangle$}} 
\begin{eqnarray}
\label{a1}
 \langle\pi^i(q)|q^a_\alpha(0)\bar q^b_\beta(0)|0\rangle&=&i(f_1\gamma_5+f_2\fmslash q
 \gamma_5)_{\alpha\beta}\left({\tau_i\over 2}\right)_{ab}\;,\\
 \label{a2}
\langle\pi^i(q)|D_\mu q^a_\alpha(0)\bar q^b_\beta(0)|0\rangle&=&\{(g_1+g_2\fmslash q)q_\mu\gamma_5
+h_1\gamma_\mu\gamma_5+h_2i\sigma_{\nu\mu}\gamma_5q^\nu\}
_{\alpha\beta}\left({\tau_i\over 2}\right)_{ab}\;,\\
\label{a3}
 \langle\pi^i(q)|D_\mu D_\nu q^a_\alpha(0)\bar q^b_\beta(0)|0\rangle&=&i\{[g_{\mu\nu}(a_1+
 a_2\fmslash q)-i\sigma_{\mu\nu}b_1+i\varepsilon_{\mu\nu\rho\sigma}\gamma^{\rho}q^{\sigma}
 \gamma_5b_2\nonumber\\&&\mbox{}+(q_\mu\gamma_\nu+q_\nu\gamma_\mu)c_1
 +(q_\mu\gamma_\nu-q_\nu\gamma_\mu)d_1\nonumber\\&&\mbox{}
 +i(q_\mu\sigma_{\lambda\nu}+q_\nu\sigma_{\lambda\mu})q^\lambda c_2
 +i(q_\mu\sigma_{\lambda\nu}-q_\nu\sigma_{\lambda\mu})q^\lambda d_2\nonumber\\&&\mbox{}
 +(e_1+e_2\fmslash q)q_\mu q_\nu]\gamma_5\}_{\alpha\beta}
 \left({\tau_i\over 2}\right)_{ab}\;;
\end{eqnarray}
\begin{eqnarray}
\label{a4}
f_1=-{1\over 2f_\pi}\langle\bar qq\rangle_0\;, \hspace{1.2cm}
f_2=-{f_\pi\over 2}\;;\hspace{1.6cm}\\
\label{a5}
g_1={1\over 4f_\pi}\langle\bar qq\rangle_0\;, \hspace{0.4cm}
g_2={f_\pi\over 4}\;,\hspace{0.4cm}  h_1=0\;, \hspace{0.4cm} 
h_2=-{1\over 12f_\pi}\langle\bar qq\rangle_0\;; 
\end{eqnarray}
\begin{eqnarray}
\label{a61}
a_1=-{m_0^2\over 16f_\pi}\langle\bar qq\rangle_0\;, \hspace{0.4cm}
a_2=-{ 5\over 36}f_\pi m_1^2\;,\hspace{0.4cm}
c_1= { 1\over 36}f_\pi m_1^2\;,\hspace{0.4cm} 
c_2=-{1\over 24f_\pi}\langle\bar qq\rangle_0\;,   
\end{eqnarray}
where $\displaystyle{\langle\pi^i(q)|\bar q\;{\frac{\tau_i}{2}}%
\;D^2\gamma_\rho\gamma_5 q|0\rangle=-if_\pi\; m_1^2\;q_\rho}$~. 
\begin{eqnarray}
 \label{a62}
  &&e_1(n\cdot q)^2={i\over 2}\langle\pi^i(q)|\bar q\;{\tau_i\over 2}\;(n\cdot D)^2
  \gamma_5 q|0\rangle\;,~~~~\mbox{for}~~~n^2=0\;,\hspace{0.5cm}\\\label{a63}
&&e_2(n\cdot q)^3={i\over 2}\langle\pi^i(q)|\bar q\;{\tau_i\over 2}\;\gamma_5\fmslash n
  (n\cdot D)^2 q|0\rangle\;,~~~~\mbox{for}~~~n^2=0\;,\hspace{0.5cm}\\
\label{a64}
   &&b_1={a_1\over 3}\;,\hspace{0.4cm}b_2=-3c_1\;,\hspace{0.4cm}d_1=0\;,\hspace{0.4cm}
   d_2=-{e_1\over 2}+{1\over 12f_\pi}\langle\bar qq\rangle_0\;.
\end{eqnarray} 
Here $f_\pi=93$ MeV. 
\vspace{1.cm}


\newpage
{\bf Figure Captions}
\vspace{2ex}
\begin{center}
\begin{minipage}{120mm}
{\sf Fig. 1.} \small{ Dependence of $\cal G$ and $\cal G'$ on the Borel parameter
$T$. The three curves  correspond, from top to bottom,  to the sum rules (\ref{sum1a}),
(\ref{sum2}) and (\ref{sum3}), respectively. }
\end{minipage}
\end{center}
\vspace{0.8cm}

\begin{table}[htbp]
\caption{ The decay widths of $D_1$, $D_2^*$, $D'_1$ and $D'_0$. The numbers in the
parenthesises are results obtained by using $e_1=-0.028$ instead of $-0.014$.}
\label{tab:width}\vspace{2mm}
\centering
\begin{minipage}{15.5truecm}
\begin{tabular}{|l|ccc|}
Transition& Calculated Width(MeV)&Observed Width(MeV)\cite{review}& Result from
\cite{dai2}\\
\hline
$\Gamma(D_1\to D^*\pi)$&$7.0 (2.8)$&&$8.0$\\
\hline
$\Gamma(D_1\to\mbox{all} )$&&$18.9^{+4.6}_{-3.5}$&\\
\hline
$\Gamma(D_2^*\to D\pi)$&$18.2 (7.3)$&&$15$\\
\hline
$\Gamma(D_2^*\to D^*\pi)$&$6.9 (2.8)$&&$6.8$\\
\hline
$\Gamma(D_2^*\to  \mbox{all})$&&$23^{+5}_{-5}$&\\
\hline
$\Gamma(D'_0\to D\pi)$&&&\\
$\Gamma(D'_1\to D^*\pi)$&$ 45 (64)$&& \\
\end{tabular}\end{minipage}
\end{table}
\vspace{3cm}

\begin{figure}[htbp]   
\begin{center}
\setlength{\unitlength}{1truecm} 
\begin{picture}(6.8,6.8)
\put(-8.0,-22)
{\includegraphics{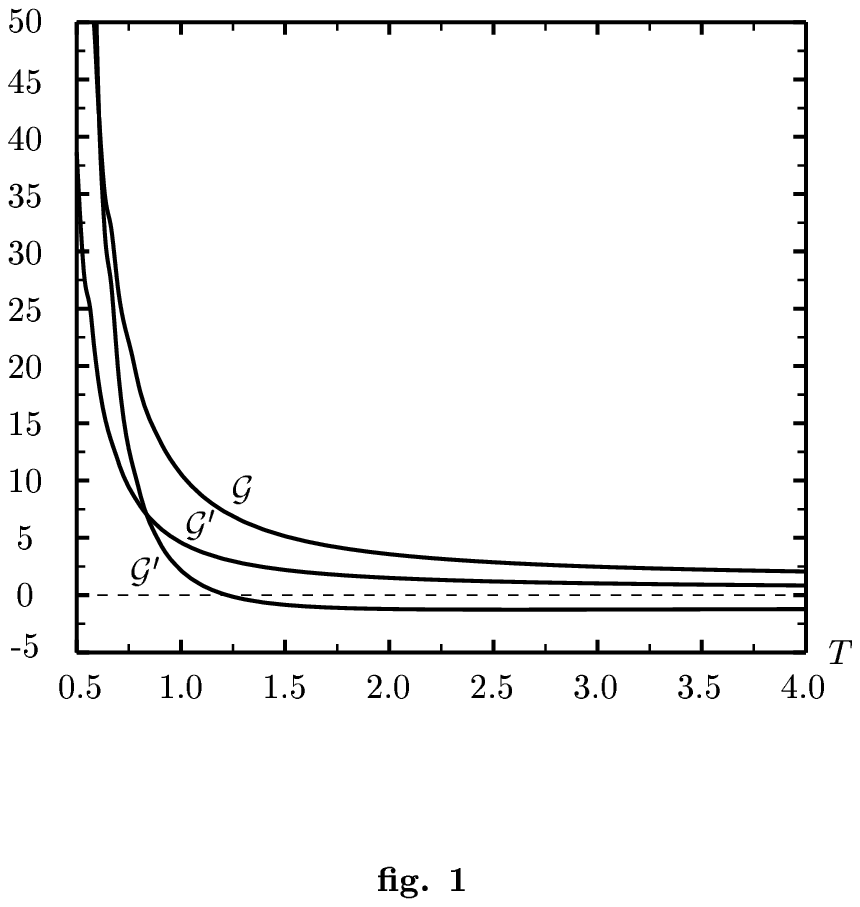}} 
\end{picture} 
\end{center} 
\vskip 2.0cm 
\protect\label{Fig.4}
\end{figure}
\end{document}